\documentclass{PoS}

\usepackage{bm}
\usepackage{mathptmx}
\usepackage{amsmath}
\usepackage{bm}
\usepackage{textcomp}

\DeclareMathAlphabet\mathbfcal{OMS}{cmsy}{b}{n}

\title{Overview on the Tunka-Rex antenna array for cosmic-ray air showers}

\ShortTitle{Tunka-Rex: Overview and Recent Results}

\author{
\speaker{F.G.~Schr\"oder$^{1}$}, P.A.~Bezyazeekov$^{2}$, N.M.~Budnev$^{2}$, D.~Chernykh$^{2}$, O.~Fedorov$^{2}$ O.A.~Gress$^{2}$, A.~Haungs$^{1}$, R.~Hiller$^{1}$\thanks{now at the University of Z\"urich}~, T.~Huege$^{1}$, Y.~Kazarina$^{2}$, M.~Kleifges$^{3}$, E.E.~Korosteleva$^{4}$, D.~Kostunin$^{1}$, O.~Kr\"omer$^{3}$, L.A.~Kuzmichev$^{4}$, V.~Lenok$^{1}$, N.~Lubsandorzhiev$^{4}$, T.~Marshalkina$^{2}$, R.R.~Mirgazov$^{2}$, R.~Monkhoev$^{2}$, E.~Osipova$^{4}$, A.~Pakhorukov$^{2}$, L.~Pankov$^{2}$, V.V.~Prosin$^{4}$, A.~Zagorodnikov$^{2}$ --
~~~~~~~~~~~~~~~~~~~~~~~~~~~~
Tunka-Rex Collaboration \\
\llap{$^1$} Institut f\"ur Kernphysik, Karlsruhe Institute of Technology (KIT), Karlsruhe, Germany\\
\llap{$^2$} Applied Physics Institute, Irkutsk State University (ISU), Irkutsk, Russia\\
\llap{$^3$} Institut f\"ur Prozessdatenverarbeitung und Elektronik, Karlsruhe Institute of Technology (KIT), Karlsruhe, Germany\\
\llap{$^4$} Skobeltsyn Institute of Nuclear Physics MSU, Moscow, Russia\\
E-mail: \email{frank.schroeder@kit.edu}       
}

\abstract{Tunka-Rex is a $3\,$km$^2$ large antenna array for cosmic-ray air showers at the TAIGA observatory (Tunka Advanced Instrument for cosmic ray physics and Gamma Astronomy) in Siberia close to Lake Baikal. 
In autumn 2016 Tunka-Rex has been extended to a total of 63 stations, each equipped with SALLA antennas for two polarization directions operating in the band of $30-80\,$MHz. 
All antenna stations are triggered by the other cosmic-ray detectors at the site, i.e., Tunka-Rex runs in coincidence with the Tunka-Grande array of scintillation detectors and during clear nights additionally in coincidence with the Tunka-133 non-imaging air-Cherenkov array. 
Compared to previous measurements, the day-time and bad-weather trigger increases the annual exposure by an order of magnitude, and the higher density of antennas is expected to increase the accuracy - in particular for the cosmic-ray mass composition. 
In addition to the status and potential of Tunka-Rex an overview of recently published results will be presented: by comparing the radio and the air-Cherenkov signals of the same air showers, the Tunka-Rex precision for the most important shower parameters has been determined. 
The direction precision is about $1^\circ$, the precision for the depth of the shower maximum, $X_\mathrm{max}$, is better than $40\,$g/cm$^2$ for high-quality events, and the energy precision is better than $15\,\%$. 
A rough estimation of the shower energy and of $X_\mathrm{max}$ is even possible with only a single antenna station when using the shower geometry measured by Tunka-133. 
Moreover, by using exactly the same external calibration source for the antennas as at the LOPES experiment we have shown that the absolute energy scale of different air-shower arrays can be compared with each other at $10\,\%$ precision. Thus, Tunka-Rex will improve the total accuracy of TAIGA for the energy as well as for the mass composition of cosmic rays of energy above $0.1\,$EeV.
}

\FullConference{35th International Cosmic Ray Conference --- ICRC2017\\
		10--20 July, 2017\\
		Bexco, Busan, Korea}

\begin{document}

\section{Introduction}
When Tunka-Rex started its operation in 2012 with 18 antennas it had two main goals. 
First, a cross-calibration with the Tunka-133 photomultiplier array using the established non-imaging air-Cherenkov technique for air showers \cite{Tunka133_NIM2014}. 
By this the precision of the radio measurements for the energy and the depth of the shower maximum, $X_\mathrm{max}$, could be determined experimentally \cite{TunkaRex_Xmax2016}. 
This goal is complementary to the comparison of radio to air-fluorescence measurements at the Pierre Auger Observatory \cite{Holt_AERA_ICRC2017}. 
It also gives additional confidence to the measurements of radio arrays whose analysis is based mainly on Monte Carlo simulations m\cite{BuitinkLOFAR_Xmax2014}. 
Second, the practical demonstration that the radio technique is cost-effective -- at least when the antennas are attached as extension to other detectors in order to improve the total accuracy. 

Meanwhile both goals have been achieved and the new main goal of Tunka-Rex is a mass-sensitive measurement of the energy spectrum between $10^{17}$ and $10^{18}\,$eV, since in this energy range the transition from galactic to extra-galactic cosmic-rays is assumed, but not yet understood. 
For this purpose the Tunka-Rex array has been extended to 63 antennas that measure in coincidence with particles and air-Cherenkov detectors (the latter only during clear nights).  
In combination with other arrays at different locations of the Earth, Tunka-Rex can search for anisotropies of different mass components and consequent differenced of the energy spectrum between the northern and southern hemispheres. 
A precise measurement of the energy spectrum can be used to study whether the second knee really exists at a few $100\,$PeV and whether it is a distinct feature from the heavy knee discovered by KASCADE-Grande \cite{KGheavyKnee2011}. 
Finally, Tunka-Rex remains a testbed for future developments of the radio technique. 
While the physics of the radio emission by air showers is understood sufficiently well \cite{HuegeReview2016, SchroederReview2016}, significant technical developments have to be done for sparse and large arrays of the next-generation, such as GRAND \cite{GRAND_ICRC2017}.

\begin{figure*}[t]
  \centering
  \includegraphics[width=0.7\linewidth]{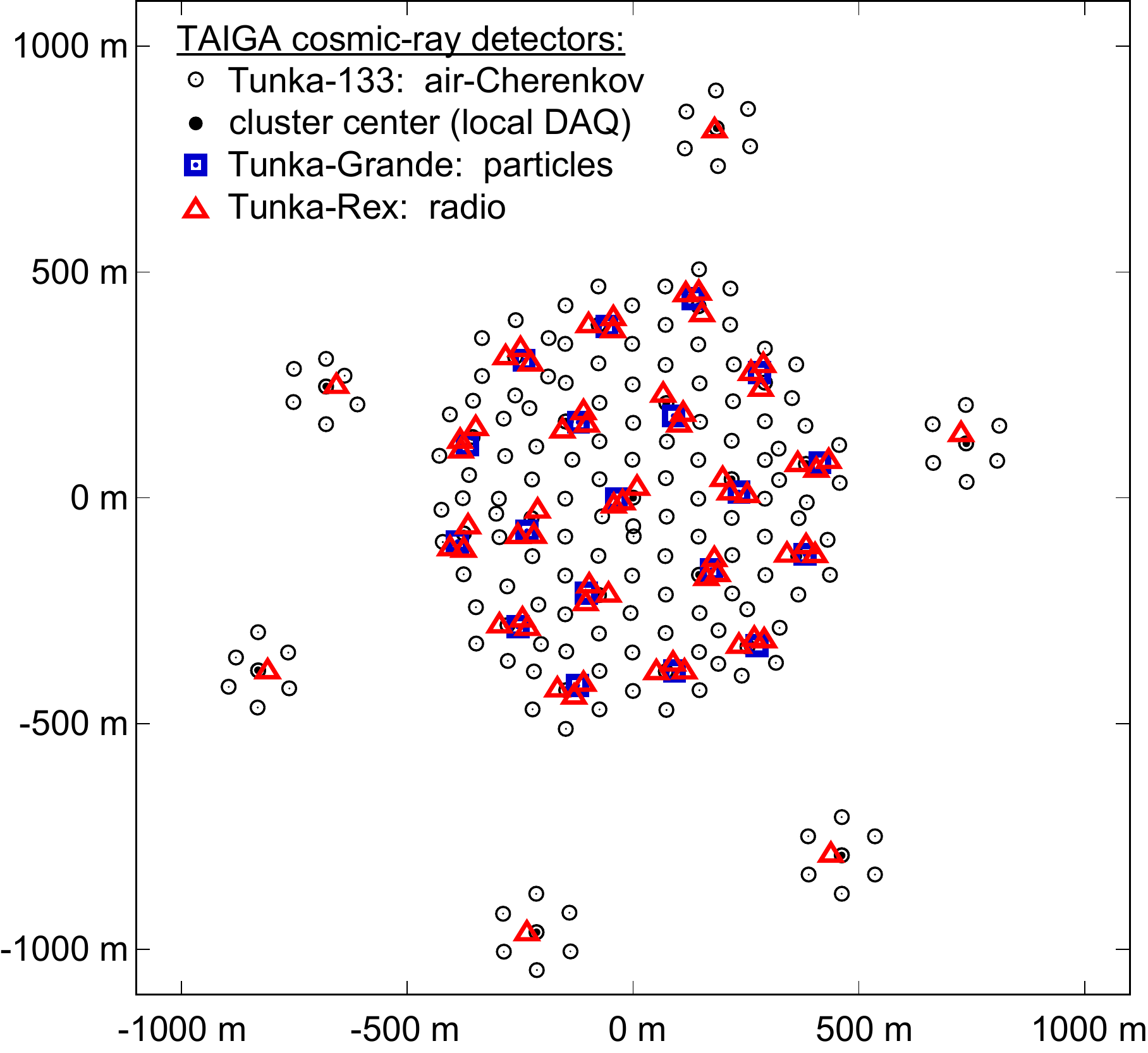}
  \caption{Map of the cosmic-ray detectors of TAIGA: the Tunka-133 air-Cherenkov array consisting of 25 clusters of seven photomultipliers each and a local data acquisition (DAQ) at each cluster center, the Tunka-Grande particle-detector array with one detector station at each of the 19 inner cluster centers, and the Tunka-Rex radio array, with 3 antenna stations at each of the 19 inner clusters triggered by both, Tunka-133 and Tunka-Grande, and one antenna stations at each outer cluster triggered by Tunka-133.}
  \label{fig_tunkaRexMap}
\end{figure*}

\begin{figure*}[t]
  \centering
  \includegraphics[width=0.99\linewidth]{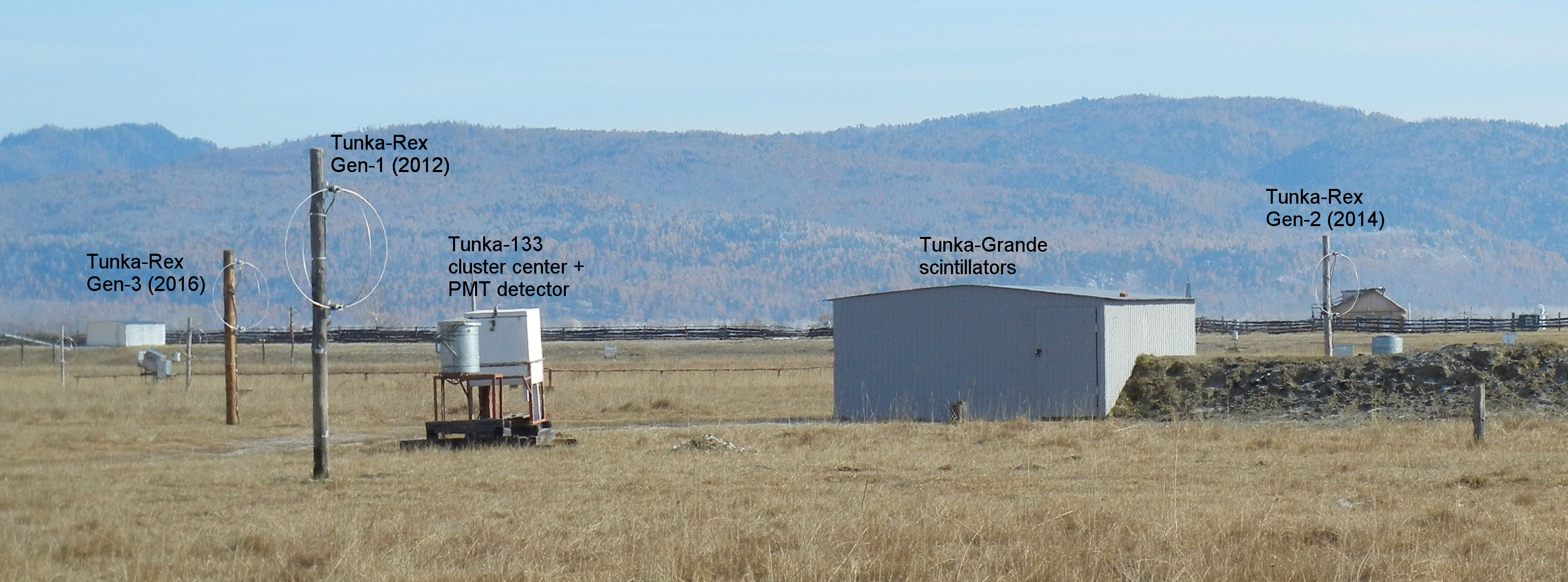}
  \caption{Photo of one cluster center with three antenna stations, a particle-detector station with surface scintillators in the gray shelter and underground scintillators in a concrete tunnel below the longish hump attached to the shelter, and a photomultiplier detector (silver cylinder) next to the local data-acquisition of this cluster (white box).}
  \label{fig_tunkaPhoto}
\end{figure*}

\section{Status of Tunka-Rex}
Since October 2016 Tunka-Rex consists of 63 antenna stations distributed over a total area of about $3\,$km$^2$ (figure \ref{fig_tunkaRexMap}) \cite{Schroeder_TunkaRex_ARENA2016}. 
Each of the 19 inner clusters of the cosmic-ray detectors of TAIGA (Tunka Advanced Instrument for cosmic ray physics and Gamma Astronomy) features 3 antenna stations (Tunka-Rex), 7 photomultiplier detectors for the air-Cherenkov light (Tunka-133), and 1 scintillator station with surface and underground particle detectors for electron and muon measurements (Tunka-Grande). 
The 6 outer clusters feature only 1 antenna station and 7 photomultiplier detectors, each.
During clear nights, when the air-Cherenkov array Tunka-133 operates it triggers all radio antennas, and during the remaining time the particle-detector array Tunka-Grande triggers all inner antennas. 
This means that for every air shower all 57 inner antennas are read out, and additionally the 6 outer antennas exclusively when the air-Cherenkov detector operates. 

Each radio station consists of a set of two perpendicularly aligned SALLAs with low-noise amplifiers integrated directly in the antenna \cite{AERAantennaPaper2012}. 
After passing a another filter-amplifier, the radio signal is digitized by the same digital data-acquisition used for the other cosmic-ray detectors of TAIGA (figure~\ref{fig_tunkaPhoto}) \cite{TAIGA_2014}. 
Using the core-position of the denser Tunka-133 as input, the radio data are analyzed by a special version of the Offline software frame work developed by the Pierre Auger Collaboration \cite{RadioOffline2011}, and afterwards compared to the Tunka-133 measurements. 
A combined analysis with the particle signals of Tunka-Grande is planned, too. 

Tunka-Rex is calibrated on an absolute scale using the same external reference source that was used first by LOPES and later by LOFAR \cite{TunkaRex_NIM_2015, 2015ApelLOPES_improvedCalibration, NellesLOFAR_calibration2015}. 
However, the antennas and the analog electronics of the subsequent signal chain are from three different production series, i.e., each of the inner clusters has one antenna of each series (figure~\ref{fig_exampleEvent}). 
While all three series use the same electronics scheme, the gain and phase response varies slightly from series to series, and the inter-calibration is not yet completed. 
Thus, at the moment most of the analyses are still based on almost 200 events detected by Tunka-Rex during the first two years of data taking with only one antenna per cluster triggered by the Tunka-133 air-Cherenkov detector.

\section{Event Reconstruction}
The reconstruction of the air-shower parameters consists of several steps described in more detail in the given references. 
First, the detector response is deconvoluted from the raw data using the phase and gain obtained by calibration measurements\footnote{ 
Lately we found that the correction for the phase responses was insufficient, i.e., delays were calculated too small and the distortion of the radio pulses due to dispersion was not accounted for properly. 
Nevertheless, the pulse distortion by the signal chain is only a second-order effect of about $2\,\%$ on the maximum pulse amplitude we use in our standard analyses. 
The necessary corrections on our previous results have been partially applied already: the figures presented here are still preliminary and we expect further changes that are small compared to other uncertainties, though. 
Together with other minor improvements, e.g., in fitting procedures, this is the main reason why the results in this proceedings slightly differ from the ones published earlier. 
}.
Then, we search for pulses with a signal-to-noise ratio $SNR > 10$ (in power) in a signal windows depending on the trigger time by Tunka-133 / Tunka-Grande \cite{TunkaRex_NIM_2015}, and correct the pulse amplitude for the average bias due to noise \cite{TunkaRex_Xmax2016}. 
The threshold is tuned such that pure noise has a pass-chance of slightly below $5\,\%$. 
Phase and gain of the antenna response depend on the arrival direction of the radio signal and are deconvoluted in an iterative process while reconstructing the direction of the air shower from the pulse arrival times in the individual antenna stations. 
At the moment we still use a plane-wave approximation for this purpose, since we need the arrival direction mainly as quality cut, where the expected difference to the more accurate hyperbolic wavefront of the order of $1^\circ$ is unimportant \cite{2014ApelLOPES_wavefront}.
Nonetheless, this simple method yields an accuracy of the arrival direction of about $1^\circ$ for showers with zenith angles $\theta < 50^\circ$, and all events for which the Tunka-Rex direction disagrees by more than $5^\circ$ from Tunka-133 are rejected as false-positive. 
This removes almost all events passing the SNR cut by chance, and the potentially remaining ones have not deteriorated our main analysis results, since we do not observe any obvious outliers. 

\begin{figure*}[t]
  \centering
  \includegraphics[width=0.99\linewidth]{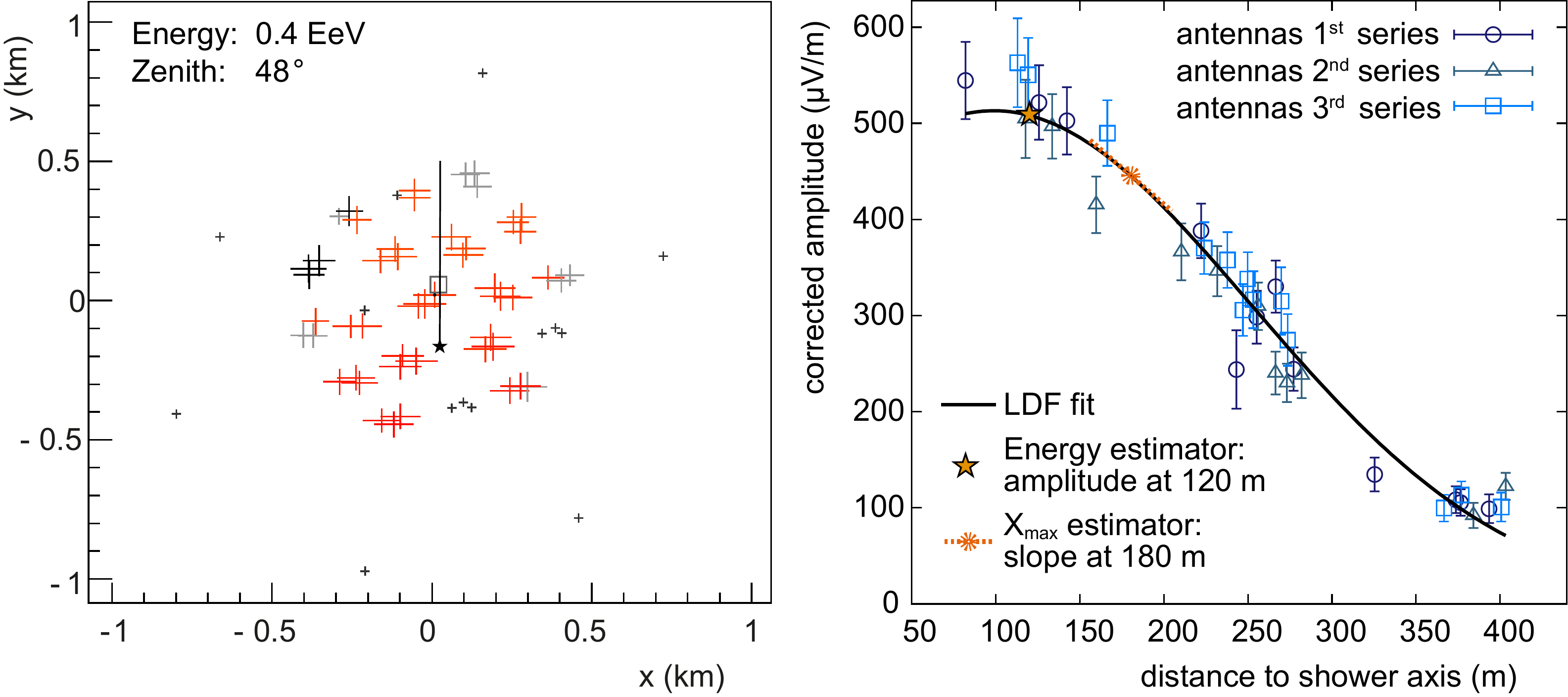}
  \caption{Example event after correction of the amplitudes in individual antennas for the geomagnetic angle and for the azimuthal asymmetry caused by the interference of the Askaryan and the geomagnetic effects. 
  Gray crosses are antenna below threshold, the small outer crosses are clusters without particle detectors that were not operating when the event was recorded.}
  \label{fig_exampleEvent}
\end{figure*}

Energy and $X_\mathrm{max}$ are reconstructed in a subsequent step from the lateral-distribution of the radio amplitude \cite{KostuninTheory2015}. 
In order to fit a simple, one dimensional lateral distribution function (LDF), we correct the amplitude in each station for the azimuthal asymmetry of the radio footprint. 
For the correction we simply assume that the strength of the Askaryan effect accounts for a constant $8.5\,\%$ of the maximum geomagnetic amplitude (i.e, more for showers that are not perpendicular to the geomagnetic field), since more complicated correction formulas did not significantly improved the subsequent accuracy for the reconstructed energy and $X_\mathrm{max}$.
Finally we fit a Gaussian LDF and use the amplitude at $120\,$m as estimator for the shower energy and the slope at $180\,$m for the reconstruction of $X_\mathrm{max}$ (see figure \ref{fig_exampleEvent} for an example event). 
All parameters, i.e., the proportionality coefficient for the energy and the function used for $X_\mathrm{max}$, have been tuned against CoREAS simulations and not to the Tunka-133 measurements we compare to.

\section{Results}
\label{section_results}
\begin{figure*}[t]
  \centering
  \includegraphics[width=0.48\linewidth]{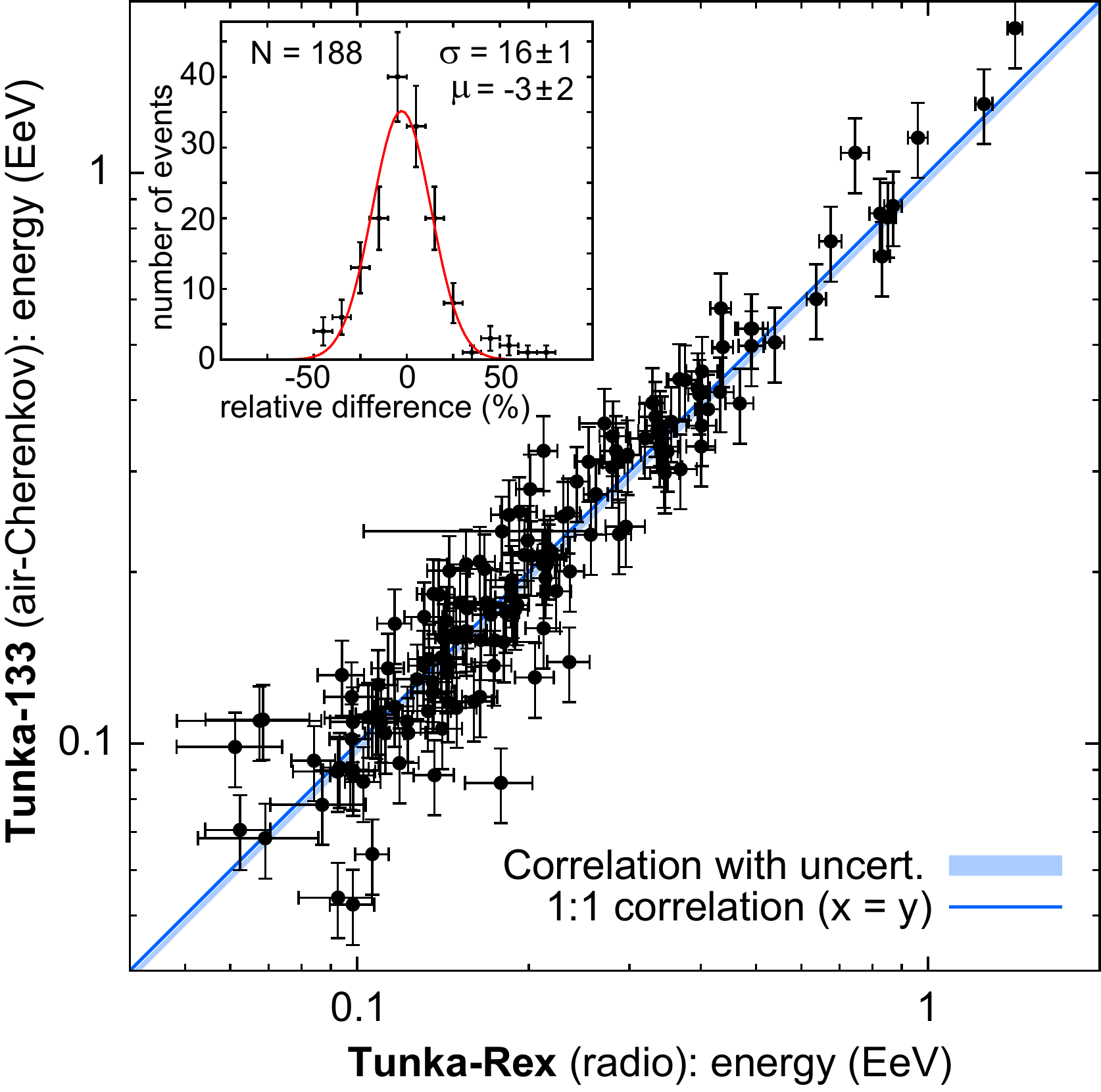}
  \hfill
  \includegraphics[width=0.50\linewidth]{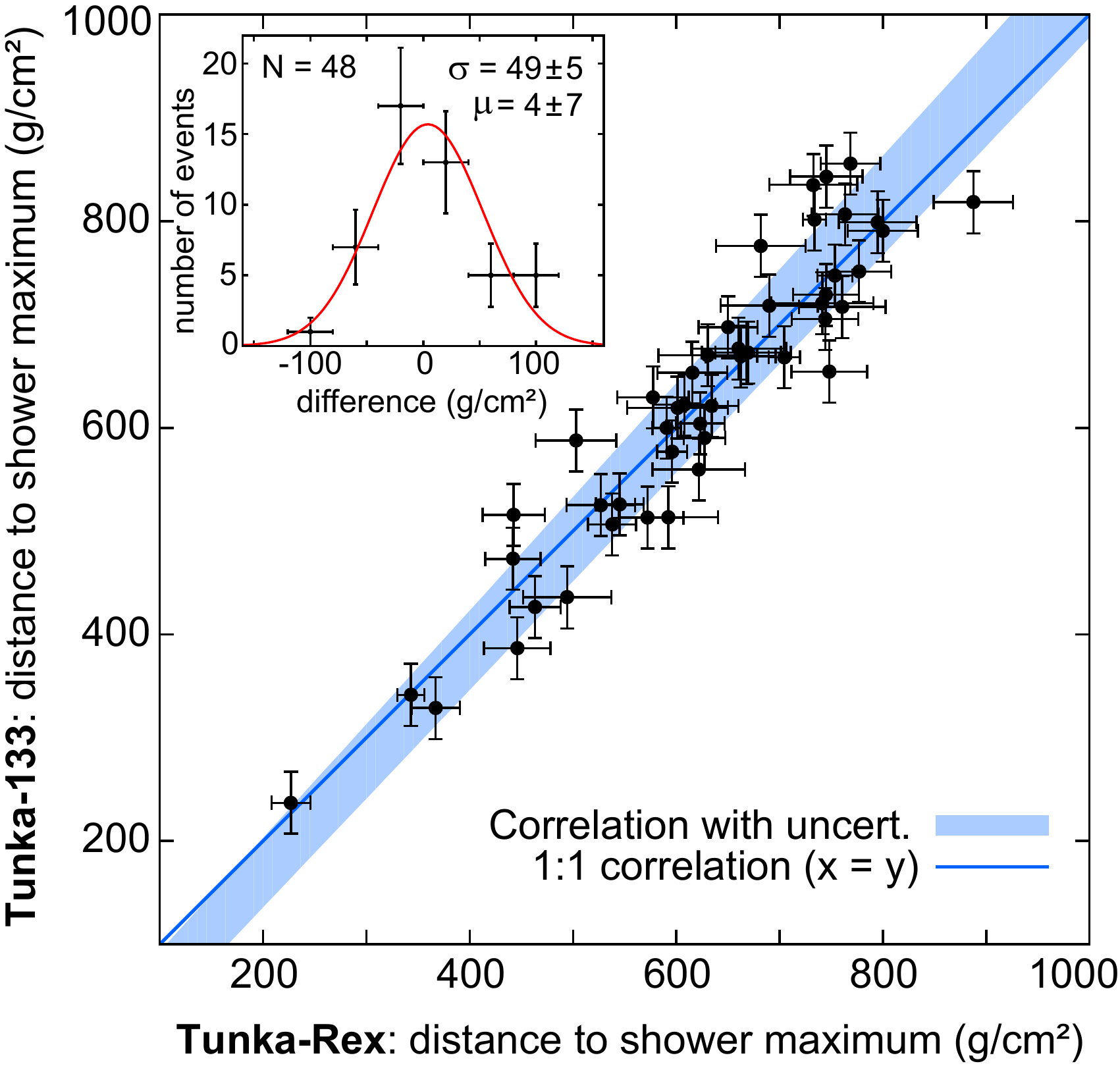}
  \caption{Correlation of the energy (left) and the distance to the shower maximum (right) between Tunka-Rex radio measurements and the Tunka-133 air-Cherenkov measurements of the same air-showers \cite{TunkaRex_Xmax2016}.}
  \label{fig_energyXmax}
\end{figure*}

Using the standard method described above we can determine the energy of every event. 
For $X_\mathrm{max}$ the uncertainties are usually too large for events with only three or four antennas above threshold which is easy to understand: 
While the axis distance of $120\,$m used for energy estimation is close to the maximum of the LDF, the amplitude is lower and closer to the noise level around the axis distance of $180\,$m used for $X_\mathrm{max}$ determination. 
Consequently, the threshold for $X_\mathrm{max}$ is about $0.2$ higher in $\lg E$ and less than one third of our events have sufficient quality to measure $X_\mathrm{max}$. 
Figure \ref{fig_energyXmax} shows the correlation of the radio and air-Cherenkov measurements of the same showers. 

As cross-check that the correlation is real and not due to any unwanted implicit tuning, we kept the Tunka-133 reconstruction of half of the events blind to the persons working on the Tunka-Rex analysis. 
The collaboration members responsible for the Tunka-133 reconstruction at first revealed only the shower axis and kept the energy and $X_\mathrm{max}$ secret until they had received the corresponding Tunka-Rex values. 
Thus, we consider the observed correlations an experimental proof that radio measurements can be used to measure not only the energy, but also $X_\mathrm{max}$, as indicated earlier by LOPES \cite{2012ApelLOPES_MTD}.
The Tunka-Rex precisions we have derived from the deviations to the Tunka-133 values are better than $15\,\%$ for the energy and about $40\,$g/cm$^2$ for $X_\mathrm{max}$. 

As next step, we aim at a lower threshold for $X_\mathrm{max}$ by further developing the computing-intensive analysis method used by LOFAR \cite{BuitinkLOFAR_Xmax2014, Kostunin_TopDownXmax_ICRC2017}.
We have already shown that the threshold can be lowered for the energy.
The energy can be reconstructed with slightly worse precision even for events with only a single antenna station above threshold when using the shower axis measured by Tunka-133 or Tunka-Grande.
This is possible because shower-to-shower fluctuations have only a small impact on the amplitude at the reference distance of $120\,$m so that we can reconstruct the energy by using the average shape of the lateral distribution \cite{Hiller_ARENA2016}. 
Moreover, we will check whether we can further improve the accuracy of our energy reconstruction by implementing ideas used by AERA \cite{AERAenergyPRL_PRD_combined2016}, and we work on a calculation of the time-, energy- and direction-dependent efficiency as basis for an energy spectrum \cite{Fedorov_TunkaRexEfficiency_ICRC2017}.

Motivated by the high accuracy of the energy reconstruction we have applied our radio measurements on an important open issue in cosmic-ray physics: the absolute energy scale of different experiments \cite{TunkaRexScale2016}. 
While the absolute scale accuracy of Tunka-Rex of $20\,\%$ is not better than the one of the host experiment Tunka-133, we were able to make a relative comparison to KASCADE-Grande with a higher accuracy of about $10\,\%$, since Tunka-Rex was calibrated by exactly the same reference source as LOPES, the radio extension of KASCADE-Grande. 
The comparison of the energy scales has been done in two different ways, both using the primary energy reconstructed by the host experiments and the radio amplitudes measured by Tunka-Rex and LOPES. 

The first method uses the ratio between the energy reconstructed by the host experiments and the radio amplitude at a common reference distance after correction for the local strengths of the geomagnetic field. 
Neglecting smaller effects, e.g., due to different observation levels and various peculiarities of the two radio arrays, different ratios between the radio amplitude and the energy directly translate into corresponding differences of the absolute energy scales of the host experiments. 
The second method is based on CoREAS simulations using the energy reconstructed by the host experiments as true input. 
Thus, the ratio between the simulated and observed radio amplitudes corresponds to the ratio of the energy scales of the host experiments. 
Since the real composition is unknown, we simulated both, a pure proton and a pure iron composition, as extreme cases with almost identical results. 
The general advantage of the second method is that it takes the specific differences between the LOPES and Tunka-Rex arrays into account because the detector simulation is included, while the advantage of the first method is that it does not depend on the theoretical models implicit in the Monte Carlo simulations. 

\begin{figure*}[t]
  \centering
  \includegraphics[width=0.39\linewidth]{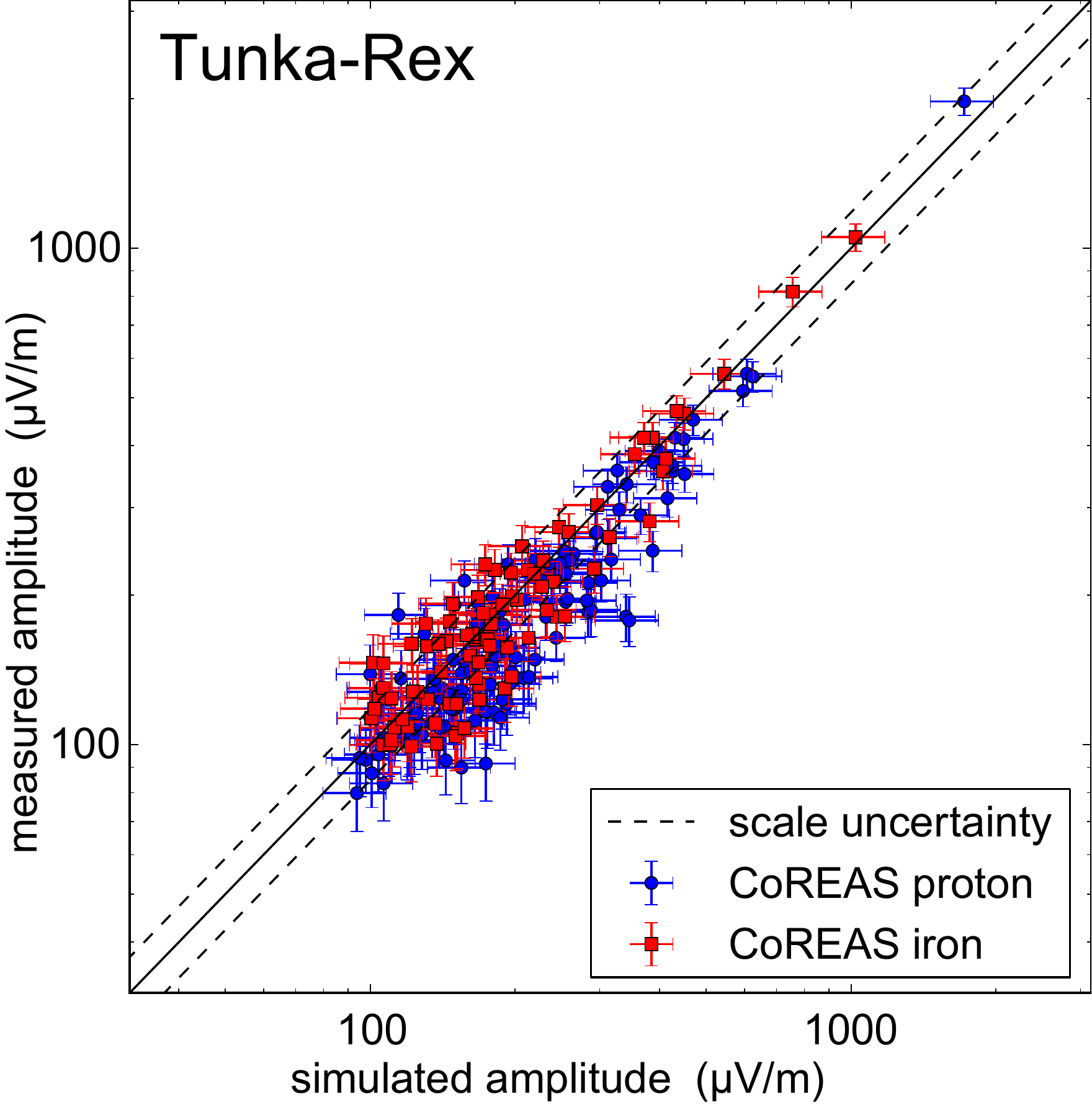}
  \hfill
  \includegraphics[width=0.59\linewidth]{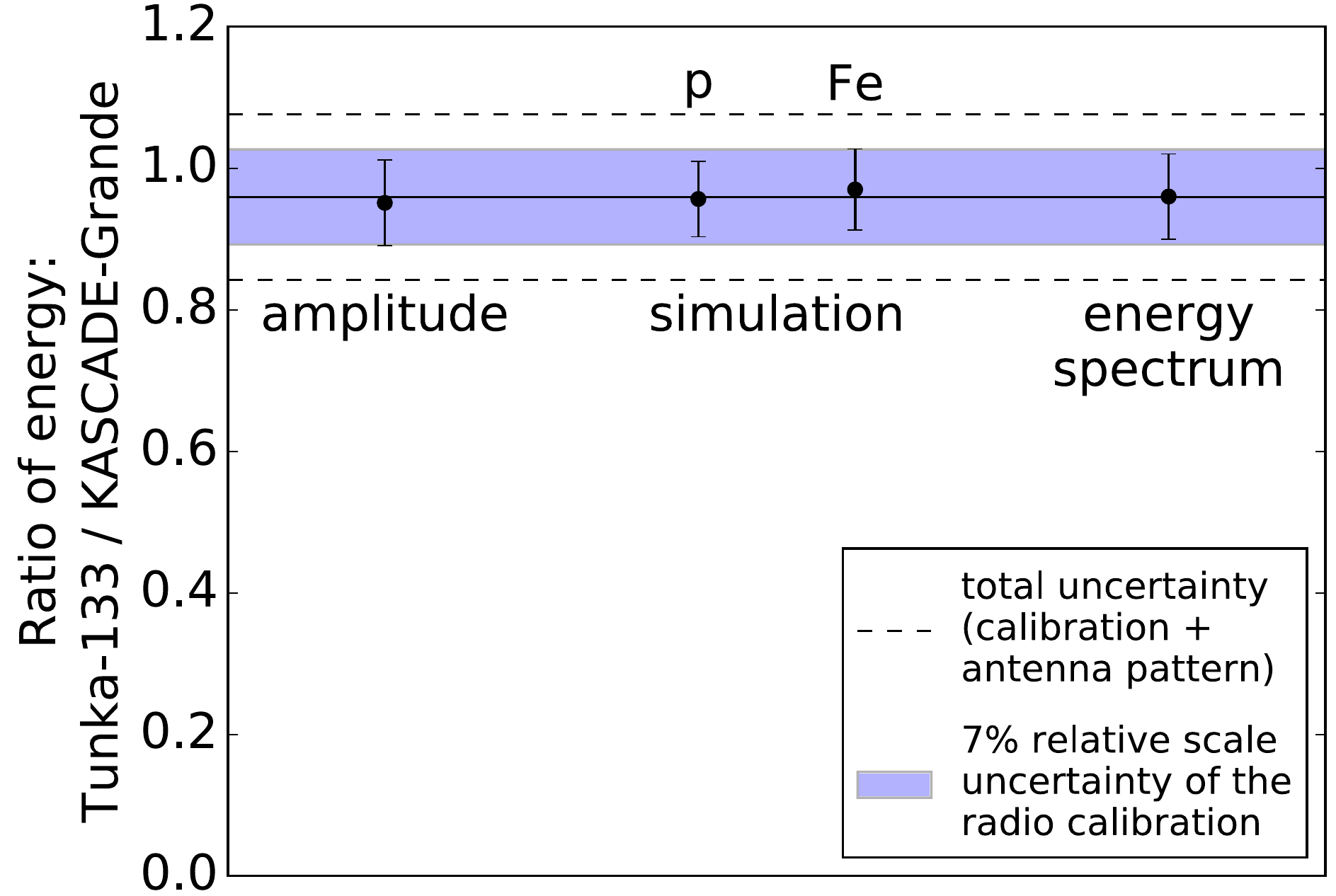}
  \caption{Left: Comparison of the radio amplitudes in individual antenna stations measured by Tunka-Rex and simulated by CoREAS using the Tunka-133 energy as input. 
  They agree within uncertainties for both extreme cases of a pure proton or a pure iron composition. 
  Right: Ratio of the Tunka-133 and KASCADE-Grande energy scales determined with different methods via their radio extension Tunka-Rex and LOPES; within a total uncertainty of about $10\,\%$ both experiments have the same absolute scale \cite{TunkaRexScale2016}.}
  \label{fig_scaleComparison}
\end{figure*}

Both methods yield the same final result within their (correlated) systematic uncertainties. 
The energy scales of Tunka-133 and KASCADE-Grande are equal within a systematic uncertainty of about $10\,\%$ (figure \ref{fig_scaleComparison}).
The small and not significant difference of the energy scales measured by the radio detectors is the same as the one obtained directly from the energy spectra of the host experiments \cite{2012ApelKGenergySpectrum, Tunka133_NIM2014}.
This is remarkable since Tunka-133 and KASCADE-Grande use completely different detection techniques, namely the measurement of air-Cherenkov light and the detection of secondary particles at ground, respectively. 

\section{Conclusion}
Tunka-Rex has shown that radio arrays can provide a cost-effective enhancement for existing air-shower arrays in order to increase their total accuracy. 
The radio measurements can compete with the established optical air-fluorescence and air-Cherenkov techniques regarding the absolute accuracy of the energy, which enables several science cases, e.g., the comparison of the absolute energy scales of different experiments. 
Moreover, radio measurements provide an measurement of $X_\mathrm{max}$ during day time. 
The $X_\mathrm{max}$ resolution of the first Tunka-Rex measurements, however, cannot yet compete with the optical techniques. 
Since these first results are based on only one antenna station per cluster, we will soon study how much the resolution improves with the current configuration of three antenna stations per cluster. 
Finally, the combination of the radio measurements with the muon measurements by Tunka-Grande will provide an additional sensitivity to the mass composition.

\clearpage

\section*{Acknowledgements}
The construction of Tunka-Rex was funded by the German Helmholtz association and the Russian Foundation for Basic Research (grant HRJRG-303).
This work has been supported by the Helmholtz Alliance for Astroparticle Physics (HAP),
by Deutsche Forschungsgemeinschaft (DFG grant SCHR 1480/1-1),
by the Russian Federation Ministry of Education and Science (projects 14.B25.31.0010, 2017-14-595-0001-003, No3.9678.2017/8.9, No3.904.2017/4.6, 3.6787.2017/7.8, 1.6790.2017/7.8), 
by the Russian Foundation for Basic Research (grants 16-02-00738, 16-32-00329, 17-02-00905),
and by grant 15-12-20022 of the Russian Science Foundation (section~\ref{section_results}).

\bibliographystyle{JHEP}
\bibliography{icrc2017}

\end{document}